\documentclass[trackchanges,twocolumn]{aastex7}
\usepackage{graphicx}
\usepackage{subcaption}
\usepackage{booktabs}
\usepackage{amsmath}
\usepackage{multirow}   
\usepackage{longtable}

\begin{document}

\title{KIC 6464285: A Solar-type Eclipsing Binary in a Hierarchical Triple System with Quasi-periodic Out-of-eclipse Modulations}

\correspondingauthor{Liying Zhu}
\email{zhuly@ynao.ac.cn}

\author[orcid=0000-0002-8320-8469,gname='Fang-Bin',sname='Meng']{Fang-Bin Meng}
\affiliation{Yunnan Observatories, Chinese Academy of Sciences (CAS), Kunming 650216, People's Republic of China}
\affiliation{University of Chinese Academy of Sciences, No.1 Yanqihu East Rd, Huairou District, Beijing, 101408, People's Republic of China}
\email{mengfangbin@ynao.ac.cn} 

\author[orcid=0000-0002-0796-7009,gname='Li-Ying', sname='Zhu']{Li-Ying Zhu} 
\affiliation{Yunnan Observatories, Chinese Academy of Sciences (CAS), Kunming 650216, People's Republic of China}
\affiliation{University of Chinese Academy of Sciences, No.1 Yanqihu East Rd, Huairou District, Beijing, 101408, People's Republic of China}
\email{zhuly@ynao.ac.cn} 

\author[0000-0002-5995-0794,gname='Sheng-Bang', sname='Qian']{Sheng-Bang Qian}
\affiliation{Department of Astronomy, School of Physics and Astronomy, Yunnan University, Kunming 650091, People’s Republic of China}
\email{qiansb@ynu.edu.cn} 

\author[orcid=0009-0000-9605-9979,gname='Ildar',sname='ASFANDIYAROV']{Ildar ASFANDIYAROV}
\affiliation{Ulugh Beg Astronomical Institute, Uzbekistan Academy of Sciences, 33, Astronomicheskaya, Tashkent, 100052, Uzbekistan}
\email{ildar@astrin.uz} 

\author[orcid=0000-0003-2999-6879,gname='Azizbek',sname='Matekov']{Azizbek Matekov}
\affiliation{Yunnan Observatories, Chinese Academy of Sciences (CAS), Kunming 650216, People's Republic of China}
\affiliation{University of Chinese Academy of Sciences, No.1 Yanqihu East Rd, Huairou District, Beijing, 101408, People's Republic of China}
\affiliation{Ulugh Beg Astronomical Institute, Uzbekistan Academy of Sciences, 33, Astronomicheskaya, Tashkent, 100052, Uzbekistan}
\email{azizbek_matekov@mail.ru} 

\author[orcid=0009-0000-2064-2551,gname='Jaqsiliq',sname='Baltamuratov']{Jaqsiliq Baltamuratov}
\affiliation{Ulugh Beg Astronomical Institute, Uzbekistan Academy of Sciences, 33, Astronomicheskaya, Tashkent, 100052, Uzbekistan}
\email{jaksilik_baltamuratov@mail.ru}

\author[gname='Wen-Ping', sname='Liao']{Wen-Ping Liao} 
\affiliation{Yunnan Observatories, Chinese Academy of Sciences (CAS), Kunming 650216, People's Republic of China}
\affiliation{University of Chinese Academy of Sciences, No.1 Yanqihu East Rd, Huairou District, Beijing, 101408, People's Republic of China}
\email{liaowp@ynao.ac.cn}

\author[orcid=0000-0001-7865-2648,gname='A-Li',sname='Luo']{A-Li Luo}
\affiliation{National Astronomical Observatories, Chinese Academy of Sciences, Beijing 100101, People's Republic of China}
\affiliation{School of Astronomy and Space Science, University of Chinese Academy of Sciences, Beijing 100049, People's Republic of China}
\affiliation{University of Chinese Academy of Sciences, Nanjing 211135, People's Republic of China}
\email{lal@nao.cas.cn} 

\author[orcid=0000-0003-0716-1029,gname='Wen',sname='Hou']{Wen Hou}
\affiliation{National Astronomical Observatories, Chinese Academy of Sciences, Beijing 100101, People's Republic of China}
\email{whou@bao.ac.cn}

\author[gname='Ping',sname='Li']{Ping Li}
\affiliation{Yunnan Observatories, Chinese Academy of Sciences (CAS), Kunming 650216, People's Republic of China}
\affiliation{University of Chinese Academy of Sciences, No.1 Yanqihu East Rd, Huairou District, Beijing, 101408, People's Republic of China}
\email{liping@ynao.ac.cn} 

\author[orcid=0000-0003-3767-6939,gname='Lin-Jia',sname='Li']{Lin-Jia Li}
\affiliation{Yunnan Observatories, Chinese Academy of Sciences (CAS), Kunming 650216, People's Republic of China}
\email{lipk@ynao.ac.cn} 

\author[gname='Er-Gang',sname='Zhang']{Er-Gang Zhao}
\affiliation{Yunnan Observatories, Chinese Academy of Sciences (CAS), Kunming 650216, People's Republic of China}
\email{zergang@ynao.ac.cn}

\author[gname='Jia-Jia',sname='He']{Jia-Jia He}
\affiliation{Yunnan Observatories, Chinese Academy of Sciences (CAS), Kunming 650216, People's Republic of China}
\email{hjj@ynao.ac.cn}

\begin{abstract}
We present the first detailed analysis of the solar-type triple system KIC 6464285. Combining long-term, high-precision photometry from Kepler, TESS, and ZTF with low-resolution spectra from LAMOST and near-infrared high-resolution spectra from SDSS/APOGEE, we performed a joint analysis of the light curves, eclipse timing variations (ETVs), and radial velocities. Spectroscopic analysis reveals the system to be triple-lined, with the inner binary’s primary being a G-type main-sequence star and a mass ratio of $0.627(7)$. Light curve modeling indicates that the inner binary is detached, with filling factors of approximately 26\% and 11\% for the primary and secondary, respectively, and a tertiary light contribution of about 27\%. ETV analysis shows a significant light-travel-time effect (LTTE), consistent with the presence of the tertiary companion, whose minimum mass is estimated as $M_{3,\rm min}=0.74(1)~M_\odot$. The light curve exhibits a pronounced O’Connell effect and quasi-periodic variations, indicative of starspot activity modulating the photometry on a $\sim$131-day timescale. Kepler observations further reveal 30 superflares, each with total energies exceeding $10^{34}$ erg. This study presents detailed observational constraints on the orbital configuration, stellar properties, and magnetic activity of KIC 6464285, providing a benchmark for studies of hierarchical triple systems.

\end{abstract}



\section{Introduction} 
Stellar multiplicity is ubiquitous in star formation and evolution, with a large fraction of solar-type stars residing in binary or higher-order multiple systems \citep{2013ARA&A..51..269D,2014AJ....147...87T, 2017ApJS..230...15M}. Recent neighborhood surveys have further refined this fraction, suggesting that approximately 17$\%$ of solar-type systems are hierarchical multiples with three or more components \citep{2021AJ....161..134H}. A typical example of such systems is the hierarchical triple system, consisting of a close inner binary and a more distant tertiary component, exhibiting a clear inner–outer orbital hierarchy. Previous statistical studies on short-period systems, such as \citet{2006AJ....131.2986P}, indicate that a significant fraction of close binaries host additional distant companions. This high occurrence rate suggests that angular momentum exchange with tertiary stars can play a key role in the formation and evolution of compact systems. Consequently, hierarchical triple systems provide an ideal laboratory to study tidal interactions, orbital evolution, angular momentum transfer, and Kozai–Lidov cycles \citep{2006A&A...450..681T, 2021Univ....7..352T}.

For hierarchical triple systems with eclipsing inner binaries, the resulting light curves (LCs)—particularly when combined with radial-velocity (RV) measurements—can precisely constrain the orbital and physical parameters of the system, such as the radii, orbital inclination, and mass ratio.
When a tertiary companion is present, the resulting light-travel time effect (LTTE) or dynamical perturbations manifest in eclipse timing variations (ETVs) \citep{1952ApJ...116..211I}. In recent years, the high-precision and long-baseline photometric data from the Kepler mission \citep{2010Sci...327..977B} and the Transiting Exoplanet Survey Satellite (TESS; \citealp{2015JATIS...1a4003R}) have greatly advanced the discovery of such systems through ETV analysis \citep{2014AJ....147...45C,2025A&A...695A.209B}. By combining LC modeling, ETV analysis, and spectroscopic RV measurements, one can obtain a coherent description of the system’s orbital architecture and stellar properties. For late-type stars, such long-baseline photometry additionally enables monitoring of starspot evolution, magnetic-activity cycles, and flare occurrences, providing direct observational diagnostics of stellar magnetic activity.

In this work, we report the discovery of KIC 6464285, a solar-type hierarchical triple system. The low-resolution spectrum from the Large Sky Area Multi-Object Fiber Spectroscopic Telescope (LAMOST; \citealt{2012RAA....12.1197C}) classifies KIC 6464285 as a G8-type star. The near-infrared spectra from the Apache Point Observatory Galactic Evolution Experiment (APOGEE; \citealt{2017AJ....154...94M}) of the Sloan Digital Sky Survey (SDSS; \citealt{2000AJ....120.1579Y}) reveal a triple-lined structure. The O–C diagram shows a significant LTTE, while LC modeling indicates the presence of a third-light contribution. Variations in the RV of the tertiary component are also detected. This study aims to constrain the orbital parameters of the third body through a combined analysis of photometry, spectroscopy, and ETVs, and to investigate the impact of starspot activity on the observed LCs. 

The structure of this paper is as follows. Section \ref{section 2} describes the photometric and spectroscopic observations. Section \ref{section ETV} presents the ETV analysis. Section \ref{section 4} provides the LC analysis, including binary modeling, starspot-induced quasi-periodic modulation, and flare activity. Section \ref{section 5} offers the discussion and conclusions.

\section{photometric and spectroscopic observations}\label{section 2}
\subsection{Photometry}
\begin{figure*}[ht!]
	\centering
	\includegraphics[width=\linewidth]{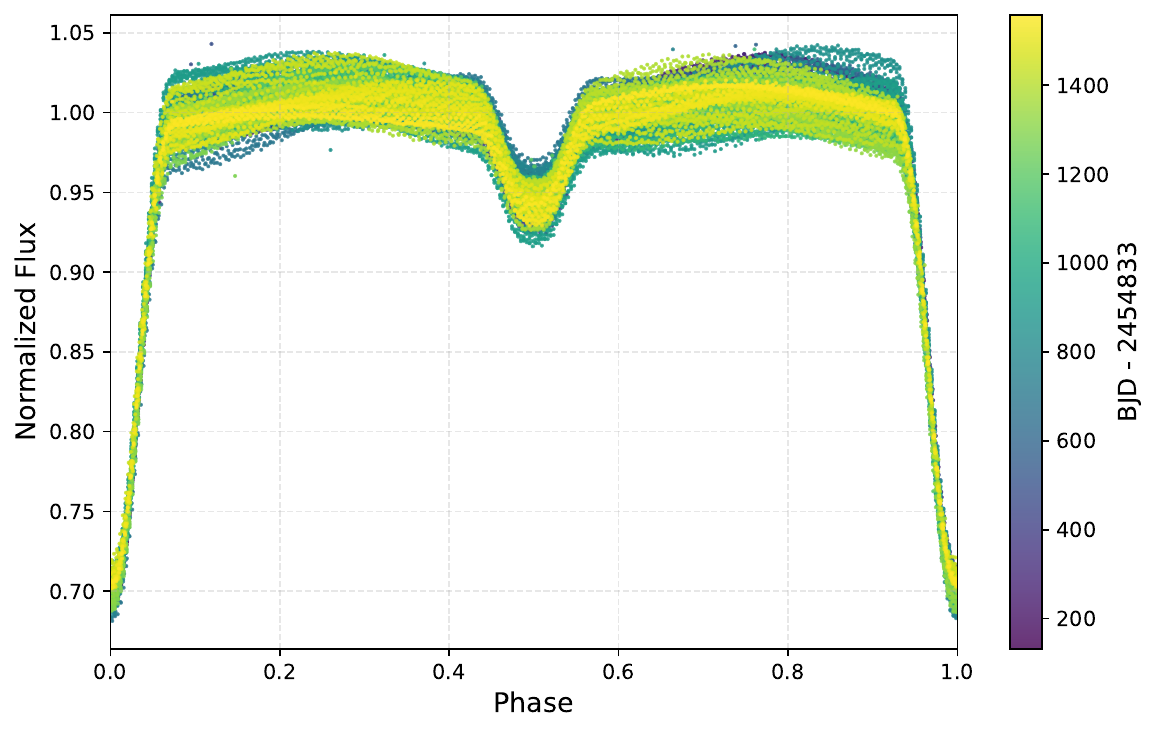}
	\caption{Kepler light curves of KIC 6464285 spanning quarters Q1–Q17. The color represents the observation time, ranging from dark blue (early epochs) to yellow (later epochs).}
	\label{Fig.1}
\end{figure*}

KIC 6464285 has been monitored by several sky surveys, including the Kepler, TESS, and the Zwicky Transient Facility (ZTF; \citealt{2019PASP..131a8002B}). Among these, the Kepler space telescope, with an aperture of 0.95 m, provided nearly continuous long-cadence photometric observations over a four-year interval (Quarters 01–17). Its superior photometric precision (e.g., $\sim 200$ ppm for $\sim 11$ mag stars at minute-level cadence) enables high-precision LC modeling and accurate determination of eclipse timings.
Figure~\ref{Fig.1} shows the Kepler LCs spanning Q1–Q17, color-coded from dark blue to yellow to indicate time evolution, normalized by the median flux of each quarter.	
In contrast, TESS employs a smaller-aperture telescope array (10.5 cm per camera) and is optimized for brighter targets to achieve comparable photometric precision. For the relatively faint target in this study ($G = 13.83$ mag), the TESS data exhibit a significantly lower signal-to-noise ratio (SNR), making them inadequate for reliably characterizing subtle out-of-eclipse variations such as the O’Connell effect and starspot-induced modulation. Similarly, the ZTF data not only have relatively low SNR, but also suffer from irregular temporal sampling, rendering them unsuitable for continuous LC analysis.
Therefore, to ensure the reliability and precision of the binary modeling, starspot characterization, and flare detection, we restrict these detailed analyses to the Kepler data. The TESS and ZTF data are instead used primarily to extract eclipse timings, thereby extending the temporal baseline of the ETV analysis.

To supplement the eclipse timing measurements obtained from survey data, we also carried out dedicated ground-based photometric observations. On 2025 June 12, observations were carried out with the 1-meter Cassegrain reflecting telescope at Yunnan Observatories (YNO), Chinese Academy of Sciences. Subsequently, on 2025 July 24–25, further observations were conducted using the eastern Zeiss-600 telescope at the Maidanak Astronomical Observatory (MAO) of the Ulugh Beg Astronomical Institute, Uzbekistan Academy of Sciences \citep{2018NatAs...2..349E}. This telescope employs a Cassegrain optical system with a 0.6-meter primary mirror and a focal length of 7.2 m, equipped with a 9600 $\times$ 6422 QHY600PH-M SBFL0 CCD camera, providing an effective field of view of approximately 17.2$\mathrm{'}\times$11.5$\mathrm{'}$ with 2$\times$2 binning and a pixel scale of 0.687$\mathrm{''}$/pixel. Johnson–Cousins BVR$_c$I$_c$ filters were used. Finally, on 2025 August 15, observations were performed with the 1.5-meter AZT-22 telescope at MAO. 
The AZT-22 adopts a Ritchey–Chrétien optical system with a focal length of 11,560 mm \citep{Artamonov1998IAUS}, equipped with an SNUCAM 4096 $\times$ 4096 CCD, providing an effective field of view of 18.1$\mathrm{'}\times$18.1$\mathrm{'}$ and a pixel scale of 0.266$\mathrm{''}$/pixel. Johnson–Bessel BVRI filters were employed to obtain multicolor, high-precision photometry \citep{2010JKAS...43...75I}. The instrumental magnitudes were transformed into the standard photometric system using the extinction coefficients and transformation equations described by \citet{Artamonov_2010ARep...54.1019A} and \citet{Azimov_2023Atmos..14.1779A}. 
All CCD images were analyzed using aperture photometry with the DAOPHOT and APPHOT packages in the Image Reduction and Analysis Facility (IRAF; \citealp{1986SPIE..627..733T}). After standard reduction, four new eclipse timings were obtained.

\subsection{Spectroscopy}\label{section 2.2}
LAMOST is a special reflecting Schmidt telescope that combines a large effective aperture ($3.6-4.9$ m) with a wide field of view $(5^{\circ})$. By employing a parallel controllable fiber positioning system, up to 4000 optical fibers can be mounted on its 1.75 m focal plane, each feeding one of 16 spectrographs, making LAMOST one of the most efficient telescopes for acquiring large numbers of stellar spectra. The low-resolution spectrum of KIC 6464285 obtained with LAMOST was processed with the LAMOST Stellar Parameter Pipeline \citep{2014IAUS..306..340W}, yielding stellar atmospheric parameters of an effective temperature $T_{\rm eff} = 5573(36)~\mathrm{K}$, a metallicity of $[\rm Fe/H] = 0.276(30)$, and a surface gravity of $\log g = 4.523(51)$.

APOGEE is a major sub-survey of SDSS-III and SDSS-IV, and represents a large-scale, high-resolution near-infrared spectroscopic survey. Its instrument design allows the simultaneous acquisition of spectra for up to 300 targets per exposure \citep{2022ApJS..259...35A}, covering the H-band (1.51-1.70 $\mu$m) at a spectral resolution of $R \approx 22{,}500$. This capability enables APOGEE to obtain high-resolution spectra even for faint companions in binary or multiple systems that are difficult to detect in the optical band. For the target KIC 6464285, a total of 19 spectra were retrieved from the SDSS DR17 database, with SNR ranging from 16 to 55.

\begin{figure}[htbp]
	\centering
	\includegraphics[width=\linewidth]{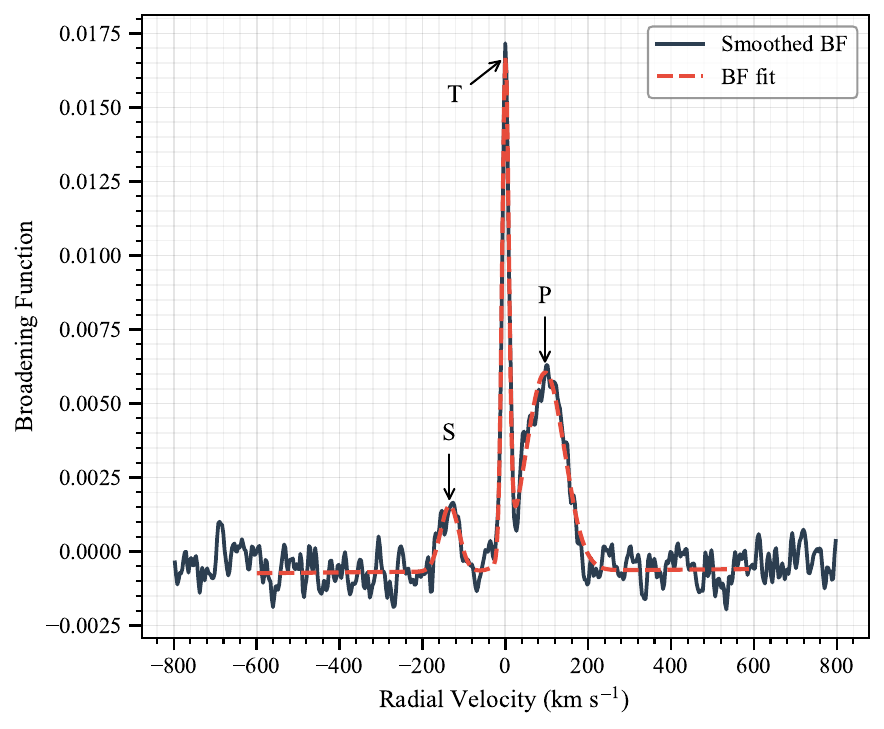}
	\vspace{2mm}
	\includegraphics[width=\linewidth]{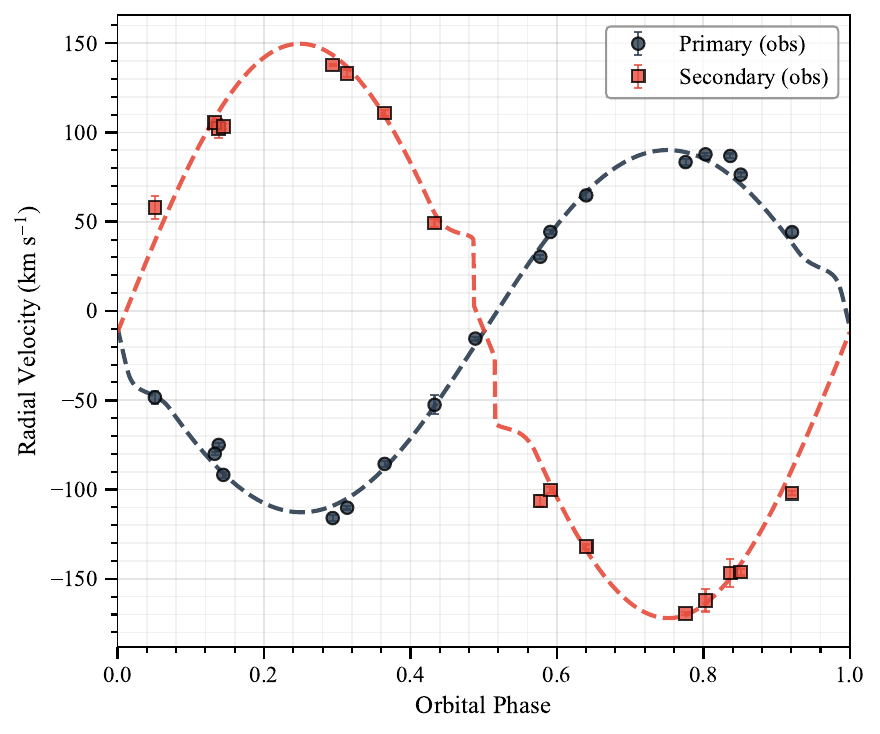}
	\vspace{2mm}
	\includegraphics[width=\linewidth]{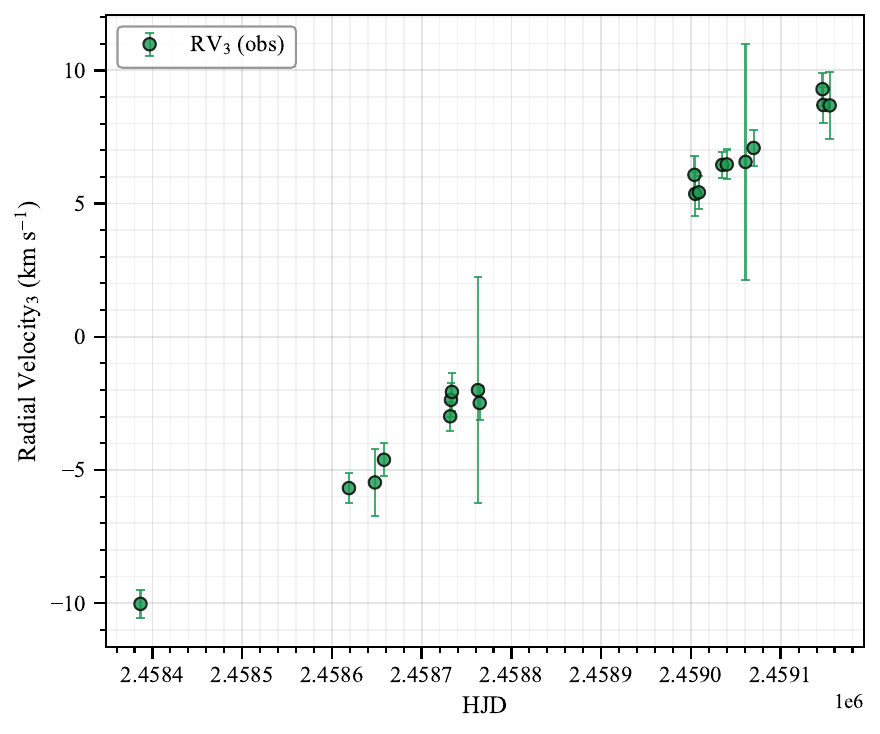}
	\caption{BF and RV analysis of KIC 6464285. 
		Top: a representative BF (blue) and its fit (red dashed line), with primary (P), secondary (S), and tertiary (T) components labeled. 
		Middle: RV curves of the primary (blue circles) and secondary (red squares) with observational error bars, overlaid with the best-fitting models. 
		Bottom: RV measurements of the tertiary component (green points).}
	\label{Fig.2}
\end{figure}

RVs were extracted using the broadening function (BF) technique developed by \citet{1999TJPh...23..271R}. The BFs were computed using PHOENIX synthetic spectra as templates \citep{2013A&A...553A...6H}, with atmospheric parameters closely matched to those derived from LAMOST low-resolution spectroscopy of the target. To reduce noise, the BFs were smoothed via Gaussian convolution in velocity space. RVs were then measured by fitting Gaussian profiles to the BF peaks. Among the 19 spectra, 16 exhibited three clear peaks, one spectrum showed two peaks, and two spectra displayed a single peak. The third component corresponds to a narrow and prominent central peak, indicating that the spectra are triple-lined. Figure~\ref{Fig.2} illustrates the spectral and RV analysis of the KIC~6464285 system. The top panel shows a representative smoothed BF together with the fitted curves, highlighting the peaks corresponding to the primary (P), secondary (S), and tertiary (T) components. The middle panel presents the measured RVs of the primary and secondary stars. These RVs were combined with the Kepler LC in a simultaneous solution using the Wilson–Devinney (W–D; \citealt{1971ApJ...166..605W,2014ApJ...780..151W}) code (see Section~\ref{Binary Modeling}). The bottom panel shows the measured RVs of the tertiary component. All measured RVs are listed in Table~\ref{tab:RV}.

\begin{deluxetable}{ccccc}
	\tablewidth{\columnwidth}
	\tablecaption{Radial velocity measurements of the three components of KIC 6464285.\label{tab:RV}}
	\tablehead{
		\colhead{JD(Hel.)}& \colhead{RV$_1$} & \colhead{RV$_2$ } & \colhead{RV$_3$ }  \\
		d&  \colhead{km s$^{-1}$} &km s$^{-1}$ &km s$^{-1}$ 
	}
	\startdata
2458386.65794 	&	$	86.86 	\pm	1.37 	$	&	$	-146.74 	\pm	8.02 	$	&	$	-10.02 	\pm	0.54 	$	\\
2458618.91622 	&	$	-75.01 	\pm	1.15 	$	&	$	102.18 	\pm	5.09 	$	&	$	-5.68 	\pm	0.57 	$	\\
2458647.84907 	&	$	-52.52 	\pm	5.28 	$	&	$	49.22 	\pm	2.06 	$	&	$	-5.47 	\pm	1.27 	$	\\
2458657.91538 	&	$	-85.64 	\pm	1.51 	$	&	$	110.80 	\pm	1.85 	$	&	$	-4.61 	\pm	0.63 	$	\\
2458731.72330 	&	$	76.38 	\pm	1.22 	$	&	$	-146.22 	\pm	1.91 	$	&	$	-2.99 	\pm	0.53 	$	\\
2458732.62609 	&	$	44.17 	\pm	2.20 	$	&	$	-102.07 	\pm	1.24 	$	&	$	-2.37 	\pm	0.63 	$	\\
2458733.64827 	&	$	-79.99 	\pm	0.91 	$	&	$	105.61 	\pm	1.22 	$	&	$	-2.07 	\pm	0.71 	$	\\
2458762.66522 	&						&						&	$	-2.00 	\pm	4.25 	$	\\
2458764.58511 	&	$	87.77 	\pm	0.93 	$	&	$	-162.12 	\pm	6.19 	$	&	$	-2.49 	\pm	0.63 	$	\\
2459003.91695 	&	$	-15.44 	\pm	1.07 	$	&						&	$	6.08 	\pm	0.72 	$	\\
2459004.83543 	&	$	30.36 	\pm	1.16 	$	&	$	-106.35 	\pm	3.16 	$	&	$	5.36 	\pm	0.84 	$	\\
2459008.83119 	&	$	-110.19 	\pm	0.99 	$	&	$	133.13 	\pm	2.14 	$	&	$	5.42 	\pm	0.61 	$	\\
2459034.96768 	&	$	-115.97 	\pm	1.76 	$	&	$	137.68 	\pm	0.71 	$	&	$	6.45 	\pm	0.50 	$	\\
2459039.90357 	&	$	-91.78 	\pm	1.52 	$	&	$	103.42 	\pm	1.41 	$	&	$	6.47 	\pm	0.56 	$	\\
2459060.83571 	&						&						&	$	6.56 	\pm	4.44 	$	\\
2459069.80814 	&	$	44.29 	\pm	1.19 	$	&	$	-100.28 	\pm	1.81 	$	&	$	7.09 	\pm	0.67 	$	\\
2459146.62178 	&	$	64.73 	\pm	2.29 	$	&	$	-131.98 	\pm	1.19 	$	&	$	9.29 	\pm	0.62 	$	\\
2459147.57984 	&	$	83.41 	\pm	1.92 	$	&	$	-169.45 	\pm	0.84 	$	&	$	8.70 	\pm	0.67 	$	\\
2459154.56134 	&	$	-48.40 	\pm	3.57 	$	&	$	58.02 	\pm	6.47 	$	&	$	8.68 	\pm	1.26 	$	\\
	\enddata
\end{deluxetable}

\section{Eclipse Timing Variations}\label{section ETV}
To investigate the orbital architecture and dynamical interactions of the system, we first analyze the ETV curve. The calculation of the O-C values relies on the initial ephemeris, expressed as:

\begin{equation}
	\operatorname{Min.} I(\mathrm{HJD})=2454964.94589428+0.8436513\ \mathrm{d} \times \mathrm{E},
\end{equation}

\noindent where the orbital period $P_0=0.8436513\ \mathrm{d}$ is adopted from the Kepler Eclipsing Binary Catalog \citep{2016AJ....151...68K}. By comparing the observed eclipse times with the predicted values from this linear ephemeris, the O-C values are obtained, which allow us to study period variations of the system. All observed times of minimum and their corresponding O–C values are listed in Table \ref{tab:eclipsing_times}. Spectroscopic observations have confirmed the presence of a third body in the system, whose gravitational influence causes the binary's barycenter to orbit the system barycenter, producing the observable LTTE in the eclipse timings. Following the model of \citet{1952ApJ...116..211I}, the LTTE term can be expressed as:

\begin{equation}
	(O-C)_{\mathrm{LTTE}}=\frac{a_{12} \sin i_3}{c}\left[\frac{\left(1-e_3^2\right) \sin \left(\nu+\omega\right)}{1+e_3 \cos \nu}+e_3 \sin \omega\right],
\end{equation}

\noindent where $a_{12} \sin i_3$ is the projected semi-major axis of the binary's barycenter around the system barycenter, $e_3, \omega$, and $\nu$ are the eccentricity, argument of periastron, and true anomaly, respectively, and $c$ is the speed of light. Considering that the initial epoch and orbital period may deviate slightly from their true values, we include linear correction terms in the O–C model. Thus, the complete theoretical O–C expression becomes

\begin{equation}
	(O-C)_{\text {model }}=(O-C)_{\text {LTTE }}+\Delta T_0+\Delta P E
\end{equation}

 \noindent where $\Delta T_0$ and  $\Delta P$ represent corrections to the initial epoch and orbital period, respectively.

\begin{figure*}[ht!]
	\centering
	\includegraphics[width=\linewidth]{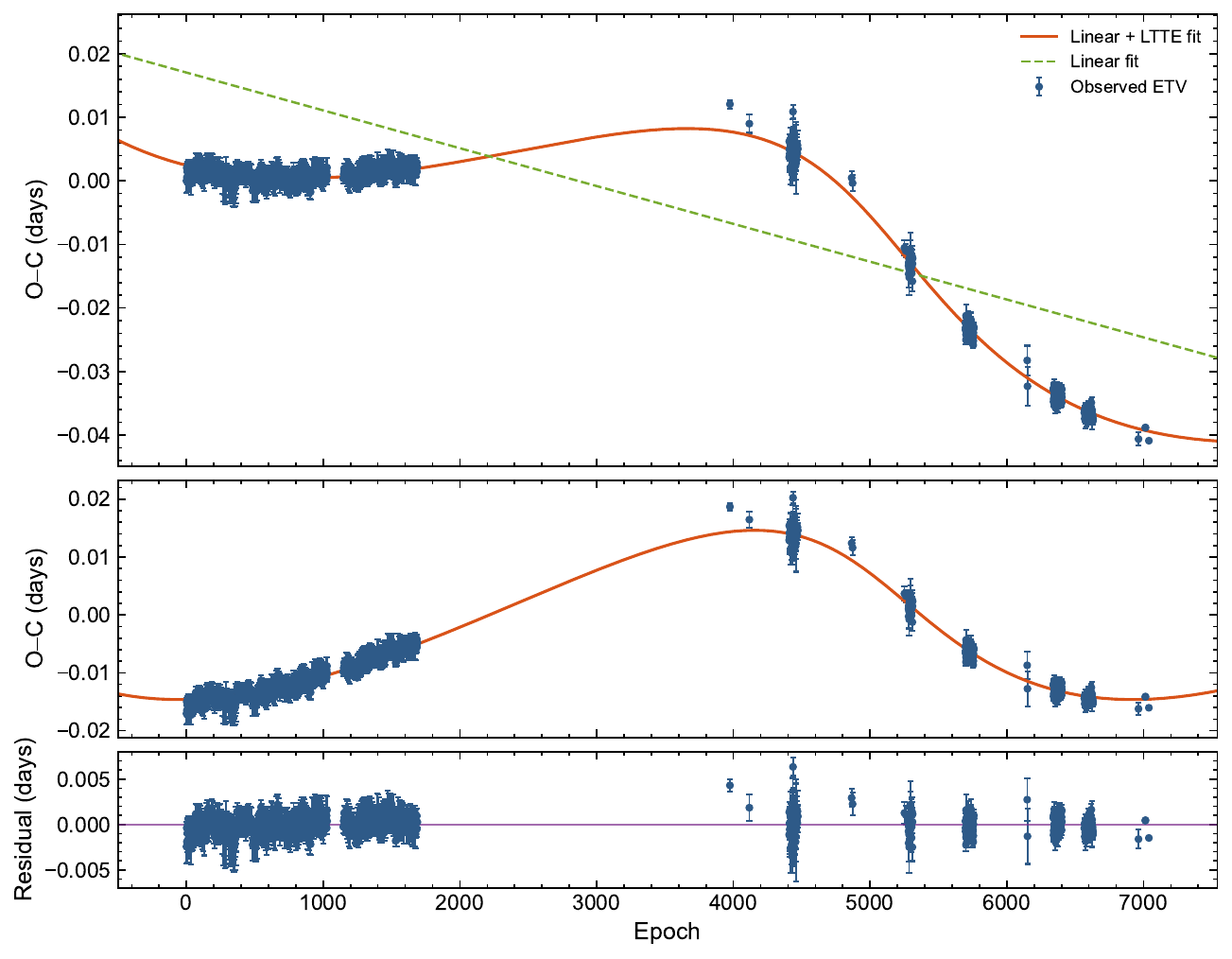}
	\caption{Observed ETVs of KIC 6464285 and the best-fit model. Top panel: observed ETVs (blue circles) with the best-fitting linear + LTTE model (orange line) and the linear trend alone (green dashed line). Middle panel: O–C diagram after subtracting the linear term, showing the LTTE. Bottom panel: residuals between the observations and the full model, with the zero line shown in purple.}
	\label{Fig.3}
\end{figure*}

\begin{figure*}[ht!]
	\centering
	\includegraphics[width=\linewidth]{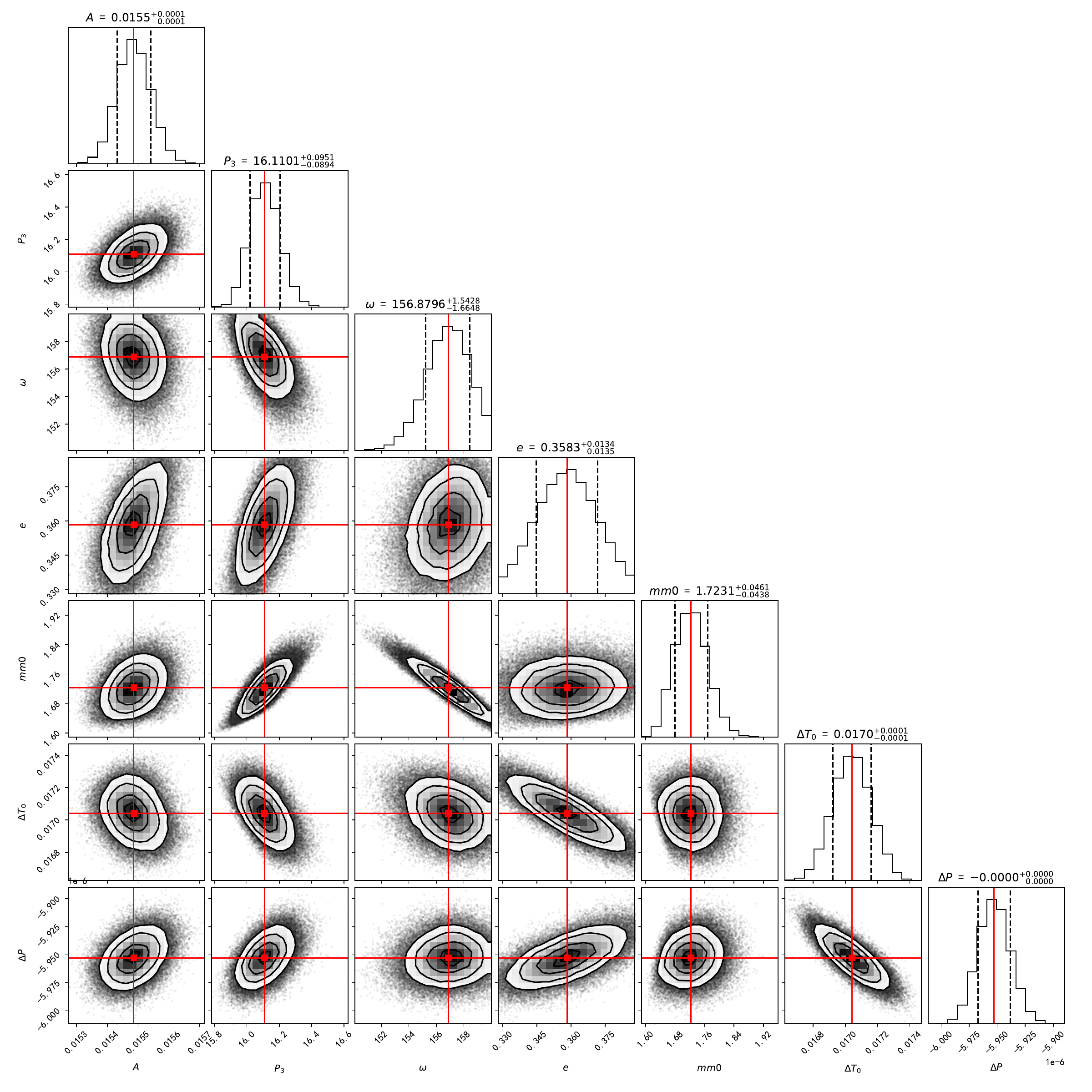}
	\caption{Corner plot showing the posterior probability distributions of the fitted parameters from the MCMC analysis of the ETV data for KIC 6464285. The parameter mm0 is defined as $mm0 = 2\pi \cdot \frac{2454964.94589428 - T_3}{365.24 \cdot P_3} $, where $T_3$ is the time of periastron passage of the third body. The units of $A, P_3, \omega, \Delta T_0$, and  $\Delta P$ are days, years, degrees, days, and days, respectively.
	}
	\label{Fig.corner}
\end{figure*}

 We performed the fitting of the O–C model using the Markov Chain Monte Carlo (MCMC) method implemented with the Python package \texttt{emcee} \citep{2013PASP..125..306F}. The best-fit O–C solution yields a third-body orbital period of $P_3 = 16.11(10)$ yr, with an LTTE amplitude of $A = 0.0155(1)$ d and an eccentricity of $e_3 = 0.36(1)$. The resulting best-fit O–C curve is shown in Figure \ref{Fig.3}, and the posterior distributions of all fitted parameters are illustrated in the corner plot (Figure \ref{Fig.corner}). The correlations between the LTTE parameters (e.g., $e_3$ and $\omega$) and the linear ephemeris terms ($\Delta P$ and $\Delta T_0$) reflect a partial degeneracy between the long-term trend and the LTTE signal, primarily caused by the limited temporal coverage of the outer orbit.

\begin{deluxetable}{ccccc}
	\tablewidth{0pt}
	\tablecaption{Times of light minima for KIC 6464285.\label{tab:eclipsing_times}}
	\tablehead{
		\colhead{JD(Hel.)} & \colhead{Error} & \colhead{E} & \colhead{O-C} & \colhead{Telescope}
	}
	\startdata
	2454964.94589 	&	0.00190 	&	0	&	0.00000 &	Kepler	\\
	2454966.63524 	&	0.00070 	&	2	&	0.00204 &	Kepler	\\
	2454967.47738 	&	0.00112 	&	3	&	0.00053 &	Kepler	\\
	2454970.00924 	&	0.00065 	&	6	&	0.00144 &	Kepler	\\
	2454970.85172 	&	0.00206 	&	7	&	0.00027 &	Kepler	\\
	2454972.54116 	&	0.00064 	&	9	&	0.00241 &	Kepler	\\
	\multicolumn{5}{c}{...} \\
	2460510.22891 	&	0.00196 	&	6573	&	-0.03697 &	ZTF	\\
	2460505.16751 	&	0.00064 	&	6567	&	-0.03647 &	ZTF	\\
	2460839.24927 	&	0.00103 	&	6963	&	-0.04063 &	YNO-1m	\\
	2460881.43360 	&	0.00011 	&	7013	&	-0.03886 &	MAO-60cm	\\
	2460882.27730 	&	0.00011 	&	7014	&	-0.03881 &	MAO-60cm	\\
	2460903.36650 	&	0.00015 	&	7039	&	-0.04089 &	MAO-1.5m	\\
	\hline
	\enddata
	\tablecomments{This table is available in its entirety in machine-readable form in the online article.}
\end{deluxetable}

\section{Light Curve Analysis}\label{section 4}

\subsection{Binary Modeling}\label{Binary Modeling}
\begin{figure*}[ht!]
	\centering
	\includegraphics[width=\linewidth]{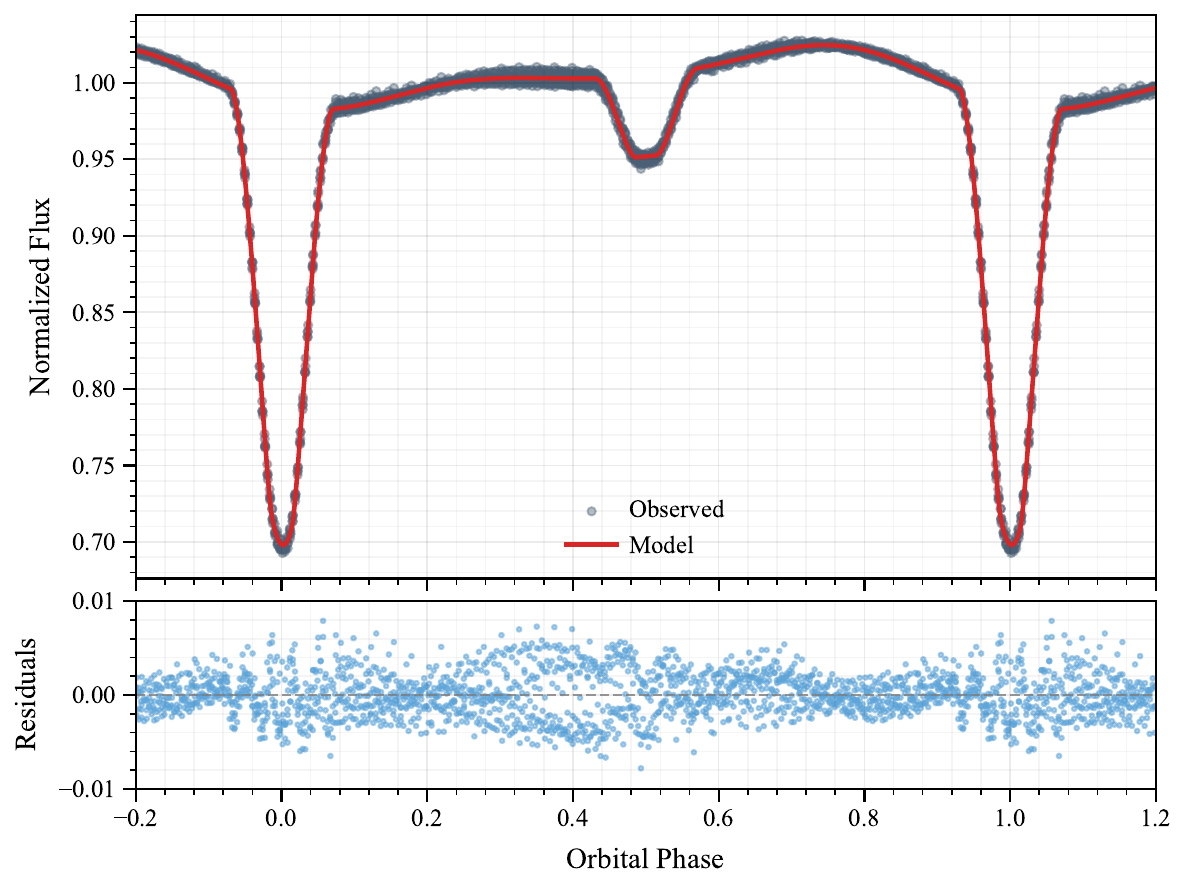}
	\caption{Kepler light curve of the binary system (BJD 2454965–2455057, Quarter 02) and the corresponding residuals. The blue-gray dots represent the observed data, and the red solid line denotes the theoretical light curve computed from the W-D model.}
	\label{Fig.WD}
\end{figure*}

To determine the physical parameters of the inner binary system, we performed a simultaneous solution of the Kepler LC and the RV curves using the 2015 version of W-D code. The Kepler data used for the modeling correspond to a relatively stable segment (BJD 2454965–2455057 in Quarter 02), containing approximately 2200 data points. In the W–D modeling, several parameters were fixed or adopted based on prior knowledge. Since the primary contributes the majority of the system’s luminosity, its effective temperature was fixed at $T_1 = 5573$ K according to the LAMOST spectroscopic results. Considering that both components of the inner binary are cool stars, the gravity-darkening coefficients were adopted as $g_1 = g_2 = 0.32$ and the bolometric albedos as $A_1 = A_2 = 0.5$. The morphology of the LC indicates that the system is a detached binary; therefore, Mode 2 of the W–D code was employed. In the simultaneous LC+RV solution, the adjustable parameters include the orbital inclination $i$, the mass ratio $q$, the systemic velocity $V_\gamma$, the semi-major axis $a$, the mean temperature of the secondary $T_2$, the monochromatic luminosity of the primary in the Kepler band, and the dimensionless surface potentials $\Omega_1$ and $\Omega_2$. To account for the presence of a tertiary component and out-of-eclipse variations, third light $l_3$ and a starspot model (including latitude, longitude, angular radius, and temperature factor) were also included in the solution. A phase shift parameter was also introduced to account for small uncertainties in the ephemeris. Figure~\ref{Fig.WD} shows the best-fitting model LC (red solid line), which reproduces the Kepler observations very well. The fitted parameters from the W-D LC+RV solution are listed in Table~\ref{tab:star_para}. Based on these parameters, the absolute parameters of the system (i.e., masses, radii, and luminosities) are derived using Kepler’s third law, and are also summarized in Table~\ref{tab:star_para}. The derived third-light flux fraction is $L_{3}/(L_{1}+L_{2}+L_{3})_{\mathrm{Kepler}} \approx 27.0 \%$. This result is independently supported by the APOGEE spectra (see Section~\ref{section 2.2} and Figure~\ref{Fig.2}), where the triple-lined BF profile shows that the tertiary component contributes an intermediate fraction of the total flux compared to the primary and secondary. A quantitative comparison based on the integrated BF profile areas (i.e., $A \times \sigma$ from Gaussian fits) indicates that the tertiary contribution is consistent with the W–D third-light fraction within $\sim 6\%$. This small difference may arise from uncertainties in the template spectra and the difference between the spectroscopic and photometric bandpasses. Overall, the agreement supports that the third-light contribution is associated with the spectroscopically detected tertiary component.

\begin{table}[ht]
	\centering
	\caption{Photometric and absolute parameters. \label{tab:star_para}}
	\begin{tabular}{l c}
		\toprule
		Parameters & Values \\
		\midrule
		$g_{1}$ & 0.32  \\
		$g_{2}$ & 0.32  \\
		$A_{1}$ & 0.5  \\
		$A_{2}$ & 0.5 \\
		$T_{1}$ (K) & 5573 \\
		$T_{2}$ (K) &  4128(3) \\
		$i~(^{\circ})$ & 87.064(82) \\
		$q~(M_{2}/M_{1})$ & 0.627(7) \\
		$L_{1}/(L_{1}+L_{2})_{\mathrm{Kepler}}$ & 0.93896(31) \\
		$L_{3}/(L_{1}+L_{2}+L_{3})_{\mathrm{Kepler}}$ & 0.2704(31) \\
		$\Omega_{1}$ & 4.4146(79) \\
		$\Omega_{1}$ & 5.0971(38) \\
		$r_{1}$ (pole) & 0.26301(24) \\
		$r_{2}$ (pole) & 0.15816(196) \\
		$r_{1}$ (point) & 0.27338(27) \\
		$r_{2}$ (point) & 0.16080(212) \\
		$r_{1}$ (side) &  0.26702(25) \\
		$r_{2}$ (side) &  0.15899(200) \\
		$r_{1}$ (back) &  0.27125(26) \\
		$r_{2}$ (back) &  0.16044(209) \\
		$f_{fill,1}$  & 0.256(10) \\
		$f_{fill,2}$  & 0.107(5) \\
		$V_\gamma~(km/s)$  & -11.165(1.206)\\
		$a~(R_{\odot})$ & 4.396(60) \\
		Latitude $\phi$ (radian) &2.86846(42)\\
		Longitude$\theta$ (radian) & 5.20743(41) \\
		Size $r_s$ (radian) & 0.62786(11) \\
		$T_s / T_*$ (radian) & 0.86273(19) \\
		\midrule
		\multicolumn{2}{c}{Absolute parameters} \\
		\midrule
		$M_{1}~(M_{\odot})$ &  0.984(41)\\
		$M_{2}~(M_{\odot})$ &  0.617(26)\\
		$R_{1}~(R_{\odot})$ &  1.173(23)\\
		$R_{2}~(R_{\odot})$ &  0.707(14)\\
		$L_{1}~(L_{\odot})$ &  1.195(56)\\
		$L_{2}~(L_{\odot})$ &  0.131(6)\\
		\bottomrule
	\end{tabular}
\end{table}

\subsection{Quasi-Periodic Variability and Photometric Modulation}
\begin{figure*}[ht!]
	\centering
	\includegraphics[width=\linewidth]{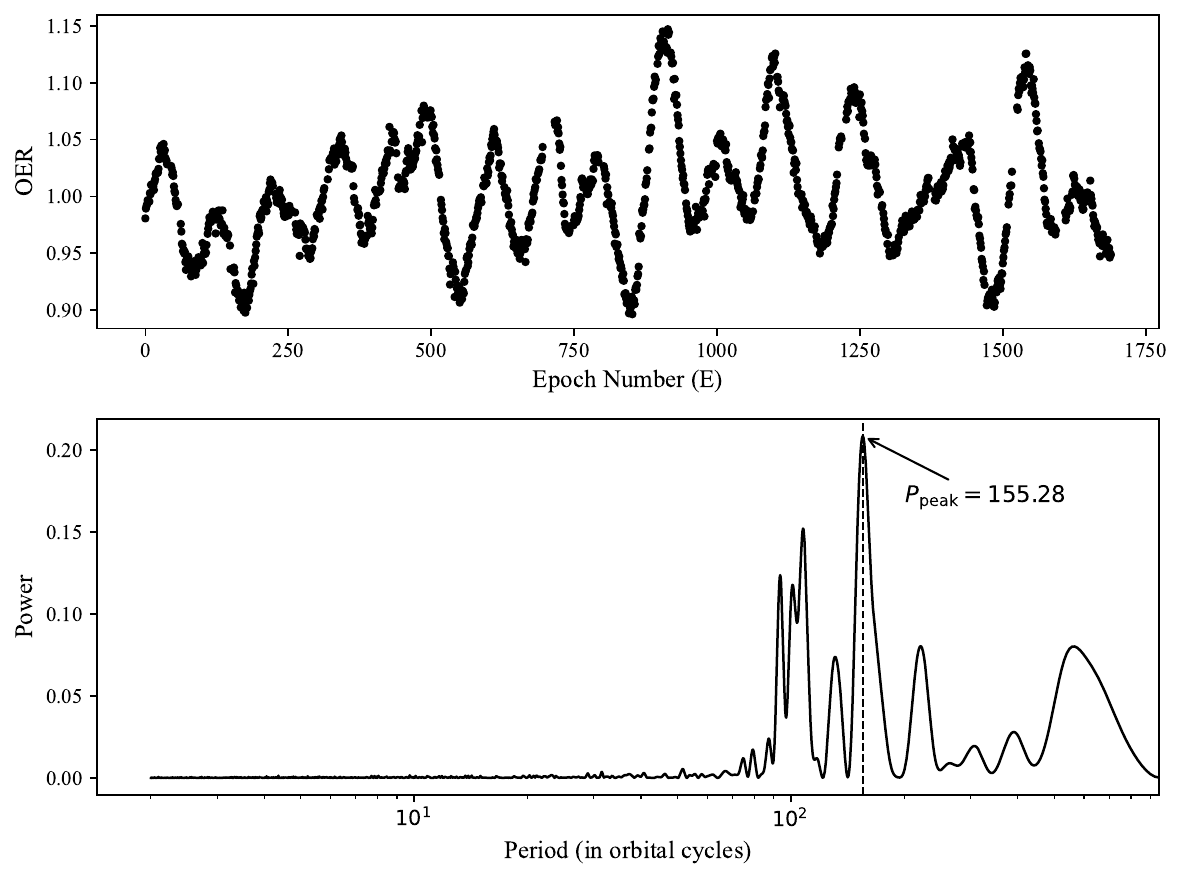}
	\caption{Upper: O'Connell Effect Ratio (OER) as a function of orbital epoch for KIC 6464285. Lower: GLS periodogram of the OER–Epoch series, showing a dominant peak at a period of 155.28 orbital cycles ($\sim$131 days).}
	\label{Fig.5-OER}
\end{figure*}

The presence of starspots can distort the out-of-eclipse light variations. As shown in Figure \ref{Fig.1}, the two maxima of the out-of-eclipse LC of KIC 6464285 are clearly unequal, exhibiting the O’Connell effect \citep{1951PRCO....2...85O}. During the observation span, the relative brightness of the orbital maxima at phases $\sim$0.25 and $\sim$0.75 alternated, giving rise to both positive and negative O’Connell effects. To quantitatively characterize this asymmetry, we adopted the O’Connell Effect Ratio (OER) proposed by \citet{1999PhDT........38M}, defined as:

\begin{equation}
	\mathrm{OER}=\frac{\sum_{i=1}^{n / 2} I_i-I_0}{\sum_{i=(n / 2)+1}^n I_i-I_0}
\end{equation}

\noindent where $I_i$ is the flux in the $i$-th phase bin, $I_0$ is the baseline flux (typically the minimum light level), and $n$ is the total number of phase bins per orbital cycle. The numerator and denominator represent the integrated flux over the left and right portions of the LC, respectively, making the OER a more comprehensive measure of the global asymmetry of the LC than a simple difference between the two maxima. The upper panel of Figure \ref{Fig.5-OER} shows the variation of OER with orbital epoch throughout the Kepler observations. The OER exhibits continuous oscillations with time. The lower panel displays the Generalized Lomb–Scargle (GLS) periodogram of the OER–Epoch series, showing a significant peak at a frequency corresponding to a modulation period of 155.28 orbital cycles, which translates to about 131 days.

\subsection{Flare Activity}
\begin{figure*}[htbp]
	\centering
	\includegraphics[width=0.49\textwidth]{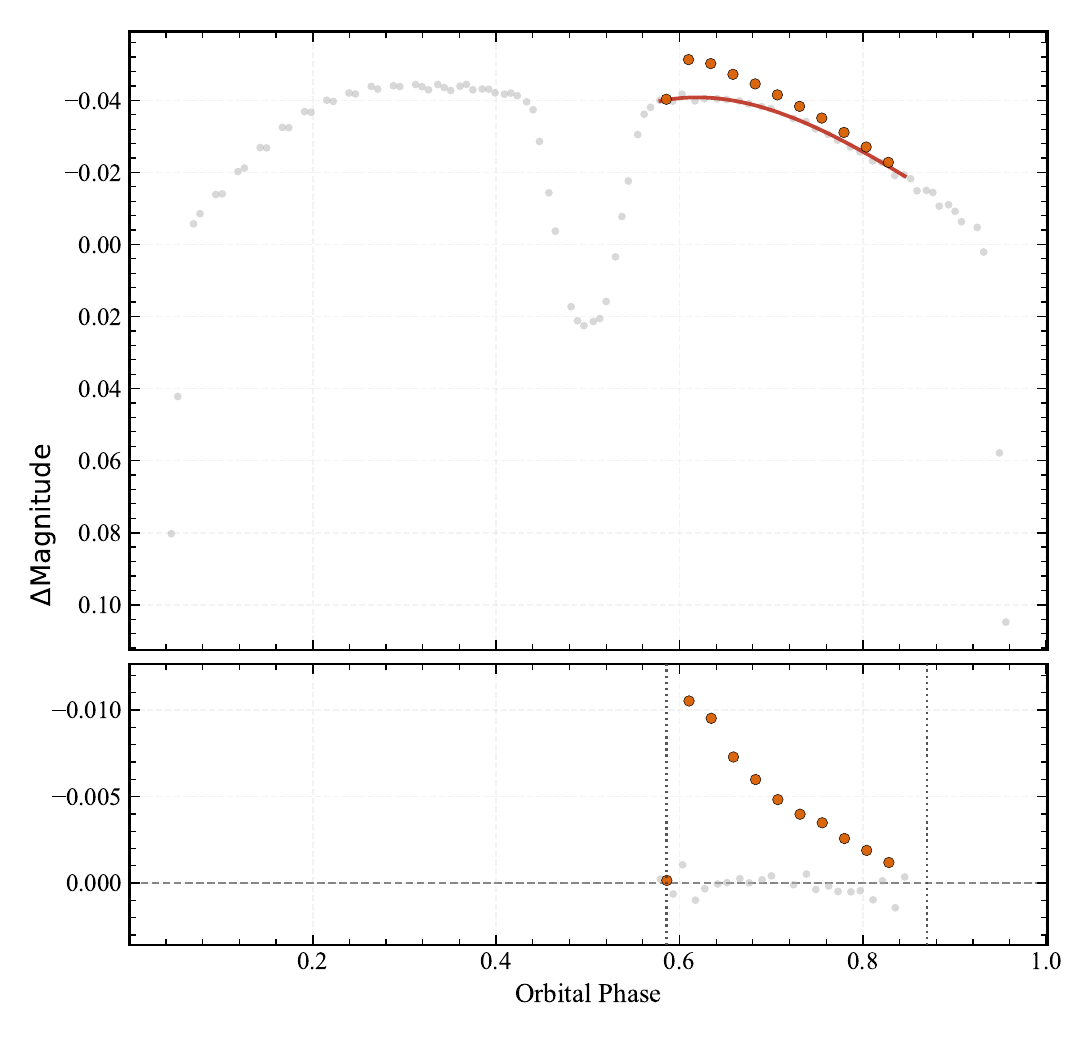}
	\hfill
	\includegraphics[width=0.49\textwidth]{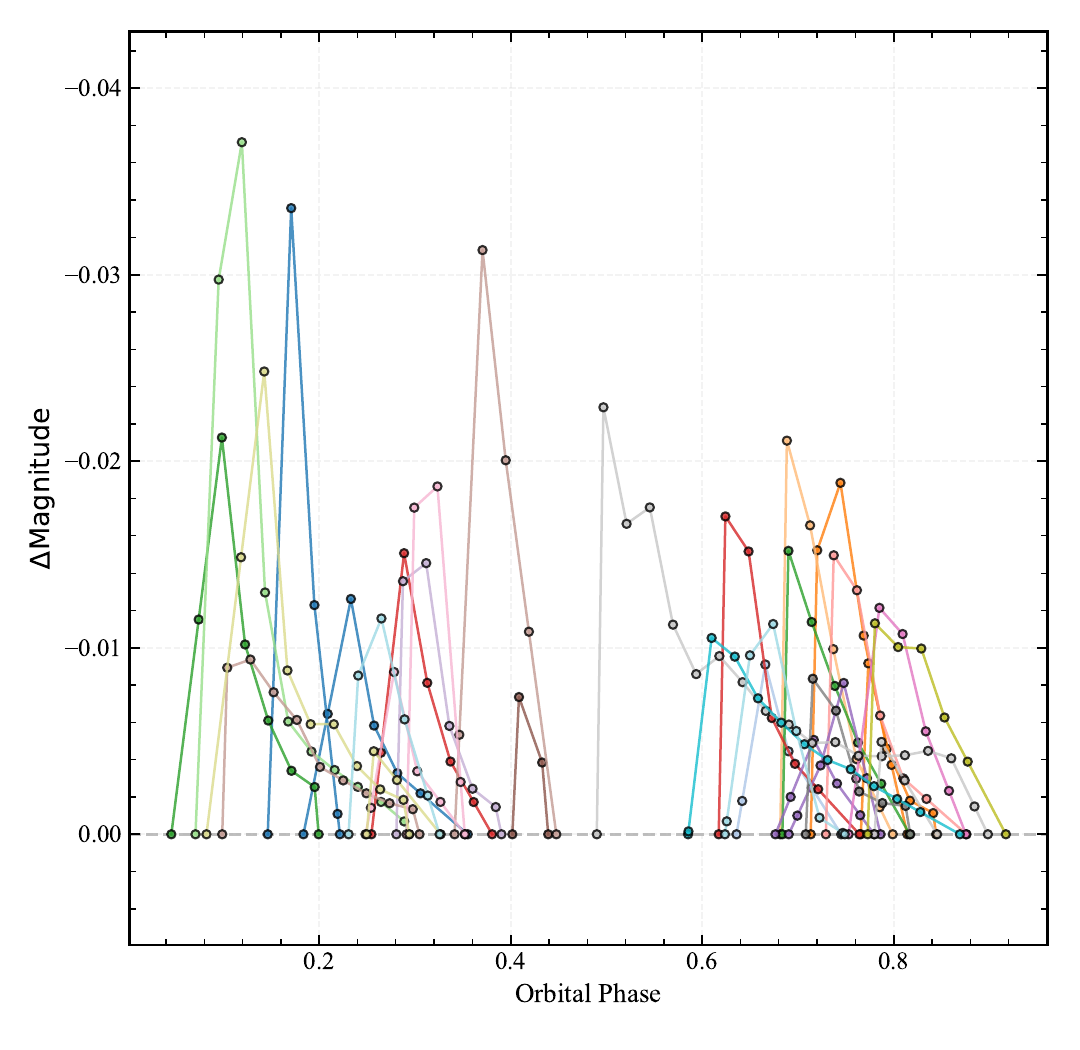}
	\caption{Identified flare events in the Kepler data. 
		Left: Example flare events in the original Kepler light-curve segments, together with the fitted local baseline used to represent the underlying binary variability.
		Right: Baseline-subtracted flare profiles folded in orbital phase, showing the cleaned morphology and peak positions of each flare.}
	\label{Fig.7-flare}
\end{figure*}

Stellar flares are typically identified in high-precision photometric data as impulsive, short-duration brightenings characterized by a rapid rise followed by a slower decay, and are widely interpreted as signatures of magnetic reconnection events (e.g., \citealt{2013ApJS..209....5S,2014ApJ...797..122D,2016ApJ...829...23D}).
In this work, flare candidates were identified using a conservative approach guided by their characteristic morphology in the LCs. Candidate events were selected as short-duration, localized brightenings that deviate from the underlying binary LC at similar orbital phases, and exhibit a rapid rise followed by a slower decay. The use of phase-folded data allows us to effectively distinguish such transient events from the periodic variability of the binary system. The flare search was initially guided by visual inspection of the Kepler LCs to identify candidate events. For each candidate, data spanning approximately two orbital cycles were extracted, ensuring that both flare-affected and flare-free segments at similar orbital phases were included. To isolate the flare signal, we constructed a local baseline representing the quiescent (flare-free) flux of the system by fitting the portions of the LC unaffected by the flare within this time window. This approach effectively uses phase-matched, flare-free data to model the underlying binary variability. The flare profile was then obtained by subtracting this baseline model from the observed flux, yielding the residual LC. As illustrated in the left panel of Figure~\ref{Fig.7-flare}, the baseline model reproduces the underlying LC, while the resulting residuals represent the isolated flare signal. The start and end times of each flare were determined by linearly interpolating the residual LC to zero flux, providing a consistent definition of flare duration. From the resulting flare profiles, we measured the start time, peak time, peak phase, end time, and duration of each event.
In total, 30 flare events were identified over the full Kepler observing span. The corresponding baseline-subtracted flare profiles are shown in the right panel of Figure~\ref{Fig.7-flare}.

The energy of each flare in the Kepler band was calculated following the method of \citet{2021AJ....161...46S}, with modifications for our triple system. The flare luminosity is expressed as:

\begin{equation}
	L_{\text {flare }}(t)=L_0(t) \times\left(10^{-0.4 \Delta M_{\text {flare }}(t)}-1\right)
\end{equation}

\noindent  where the quiescent luminosity of the system $L_0(t)$ is

\begin{equation}
	L_0(t)=\left(L_1+L_2+L_3\right) \times 10^{-0.4 \Delta M_{\mathrm{binary}}(t)},
\end{equation}

\noindent
	where $L_1$ and $L_2$ are the bolometric luminosities of the eclipsing components (see Table~\ref{tab:star_para}), and $L_3$ is estimated from the tertiary contribution using the third-light fraction derived from the W-D solution (see Section~\ref{Binary Modeling}). We note that this treatment approximates the tertiary bolometric contribution using its relative flux in the Kepler band, which introduces a systematic uncertainty of order several percent in the adopted quiescent luminosity. Since the flare energy scales linearly with $L_0(t)$, the same level of uncertainty propagates into the derived flare energies. We implicitly assume that the stellar and flare spectra are identical, and that optical variations caused by the O'Connell effect are negligible. The total flare energy is then obtained by integrating over the flare duration:

\begin{equation}
E_{\text {flare }}=\int_{\text {flare }} L_{\text {flare }}(t) d t.
\end{equation}

\noindent The total energies of the 30 detected flares all exceed $10^{34}$ erg, indicating that they are superflares. One of the events occurs near orbital phase 0.5, implying that this flare was most likely produced on the primary component. The detailed properties of all flares, including their measured parameters and derived energies, are summarized in Table~\ref{tab:flare}.

\begin{table*}
	\centering
	\caption{Parameters of the 30 flare events detected in the Kepler data. \label{tab:flare}}
	\begin{tabular}{cccccccc}
		\toprule
		No. & Start Time & Peak Time & Peak Phase & End Time & Duration & Energy & N$_{\mathrm{obs}}$ \\
		& (KJD) & (KJD) &  & (KJD) & (min) & (erg) &  \\
		\midrule
		1	&	214.781 	&	214.822 	&	0.233 	&	214.925 	&	208.471 	&	3.61E+35	&	7	\\
		2	&	216.436 	&	216.457 	&	0.171 	&	216.500 	&	91.468 	&	5.34E+35	&	5	\\
		3	&	300.371 	&	300.396 	&	0.666 	&	300.463 	&	132.884 	&	1.98E+35	&	6	\\
		4	&	311.404 	&	311.430 	&	0.744 	&	311.514 	&	159.081 	&	5.29E+35	&	8	\\
		5	&	366.285 	&	366.292 	&	0.774 	&	366.326 	&	59.023 	&	1.05E+35	&	4	\\
		6	&	394.054 	&	394.060 	&	0.688 	&	394.154 	&	142.716 	&	5.17E+35	&	7	\\
		7	&	471.134 	&	471.178 	&	0.098 	&	471.264 	&	186.927 	&	6.19E+35	&	8	\\
		8	&	473.359 	&	473.365 	&	0.690 	&	473.472 	&	161.515 	&	4.20E+35	&	7	\\
		9	&	492.246 	&	492.287 	&	0.119 	&	492.434 	&	270.949 	&	1.11E+36	&	11	\\
		10	&	557.362 	&	557.391 	&	0.289 	&	557.469 	&	153.136 	&	3.61E+35	&	7	\\
		11	&	561.043 	&	561.049 	&	0.624 	&	561.167 	&	178.461 	&	4.49E+35	&	7	\\
		12	&	591.509 	&	591.516 	&	0.737 	&	591.633 	&	178.405 	&	4.00E+35	&	7	\\
		13	&	674.998 	&	675.046 	&	0.748 	&	675.079 	&	116.523 	&	1.69E+35	&	6	\\
		14	&	675.830 	&	675.863 	&	0.717 	&	675.916 	&	125.233 	&	1.17E+35	&	6	\\
		15	&	744.675 	&	744.701 	&	0.312 	&	744.768 	&	133.334 	&	3.69E+35	&	7	\\
		16	&	824.080 	&	824.086 	&	0.409 	&	824.112 	&	45.518 	&	8.04E+34	&	4	\\
		17	&	861.150 	&	861.175 	&	0.371 	&	861.240 	&	129.096 	&	7.55E+35	&	6	\\
		18	&	925.907 	&	925.932 	&	0.128 	&	926.081 	&	250.332 	&	4.50E+35	&	11	\\
		19	&	946.706 	&	946.733 	&	0.785 	&	946.810 	&	148.766 	&	3.68E+35	&	7	\\
		20	&	975.845 	&	975.872 	&	0.323 	&	975.896 	&	73.807 	&	3.65E+35	&	5	\\
		21	&	1022.209 	&	1022.234 	&	0.278 	&	1022.297 	&	126.069 	&	1.68E+35	&	6	\\
		22	&	1042.845 	&	1042.851 	&	0.716 	&	1042.937 	&	132.576 	&	1.93E+35	&	7	\\
		23	&	1186.082 	&	1186.088 	&	0.497 	&	1186.426 	&	496.207 	&	1.46E+36	&	19	\\
		24	&	1209.949 	&	1209.955 	&	0.787 	&	1210.004 	&	79.514 	&	7.60E+34	&	4	\\
		25	&	1252.125 	&	1252.132 	&	0.780 	&	1252.247 	&	175.362 	&	4.41E+35	&	7	\\
		26	&	1318.333 	&	1318.339 	&	0.257 	&	1318.398 	&	93.845 	&	8.07E+34	&	4	\\
		27	&	1323.253 	&	1323.304 	&	0.143 	&	1323.432 	&	257.314 	&	8.07E+35	&	10	\\
		28	&	1410.574 	&	1410.595 	&	0.610 	&	1410.813 	&	344.856 	&	5.87E+35	&	13	\\
		29	&	1455.320 	&	1455.363 	&	0.674 	&	1455.426 	&	151.981 	&	3.15E+35	&	8	\\
		30	&	1501.389 	&	1501.418 	&	0.265 	&	1501.470 	&	115.460 	&	2.89E+35	&	6	\\
		\bottomrule
	\end{tabular}
\tablecomments{KJD is defined as BJD$-$2454833.}
\end{table*}

\section{DISCUSSION AND CONCLUSIONS}\label{section 5}

We present a comprehensive photometric and spectroscopic analysis of KIC 6464285. Using SDSS–APOGEE near-infrared spectra, we derived the RVs of the inner binary and the tertiary component, confirming that the system is a spectroscopic triple consisting of an eclipsing binary and a widely orbiting tertiary companion.

By combining the RV curves with the Kepler LC in a simultaneous W–D solution, we determined the mass ratio of the inner binary to be $q=0.627(7)$ and derived a self-consistent set of system parameters. The inner pair is a detached binary, with filling factors of about 26\% for the primary and 11\% for the secondary. A third-light contribution of $\sim 27\%$ is required to reproduce the Kepler photometry, consistent with the tertiary component revealed by the BF profiles. 
Using these results, we derived the absolute parameters of the system (see Table~\ref{tab:star_para}). The primary is slightly evolved from the main sequence, while the secondary remains a main-sequence star.

\begin{figure*}[ht!]
	\centering
	\includegraphics[width=\linewidth]{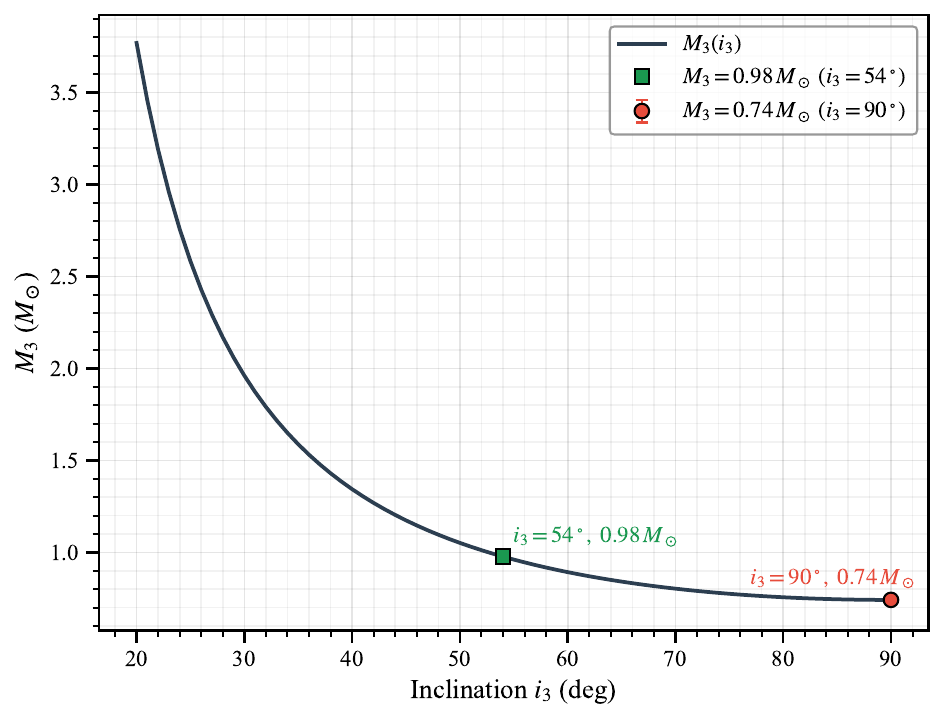}
	\caption{Tertiary mass as a function of the orbital inclination $i_3$.}
	\label{Fig.8-M3i3}
\end{figure*}

The O–C analysis was used to investigate variations in the orbital period (Figure~\ref{Fig.3}), revealing pronounced fluctuations. The continuous nature of the O–C variations, together with the triple-lined spectroscopic features, can be naturally explained by the LTTE induced by the tertiary star. Using the mass function:

\begin{equation}
	f(m)=\frac{\left(M_3 \sin i_3\right)^3}{\left(M_1+M_2+M_3\right)^2}=\frac{4 \pi^2}{G} \frac{\left(a_{12} \sin i_3\right)^3}{P_3^2},
\end{equation}

\noindent where $a_{12} \sin i_3=A \cdot c$ and $A$ is the LTTE amplitude from the O–C diagram, we obtain a mass function of $f(m) = 0.074(1)\,M_{\odot}$. The tertiary mass as a function of orbital inclination $i_3$ is shown in Figure~\ref{Fig.8-M3i3}. The minimum mass is $M_{3,\mathrm{min}} = 0.74(1)\,M_{\odot}$ at $i_3=90^{\circ}$. The inclination $i_3$ cannot be constrained from the ETV analysis alone, as only the projected quantity $a_{12}\sin i_3$ is measurable. As an illustrative case, if $i_3 \approx 54^{\circ}$, the inferred tertiary mass becomes comparable to that of the primary component, suggesting that the inclination is likely within the range $54^{\circ} \lesssim i_3 \leq 90^{\circ}$. In addition to the LTTE signal, the tertiary component also exhibits noticeable RV variations in the APOGEE spectra, further confirming its dynamical influence on the system. However, the number of available RV measurements is limited and the phase coverage is sparse, which prevents a stringent constraint on the outer orbit. Additional high-resolution spectroscopic monitoring will be required to fully characterize the tertiary orbit.

Figure \ref{Fig.1} exhibits a pronounced O’Connell effect in the system, likely caused by stellar magnetic activity or surface starspots, which is also supported by the LC modeling. To quantify the asymmetry of the light maxima, we analyzed the system using the OER metric. The OER variations reveal a rapid quasi-periodic modulation with a timescale of approximately 131 days. This behavior is analogous to the starspot migration reported by Pan \citep{2023AJ....165..247P}, which was inferred from differences between light maxima. Specifically, starspots on the stellar surface drift forward relative to the binary orbit due to differential rotation, producing periodic oscillations in the LC, with a full longitudinal rotation of the spots taking roughly 131 days.

In addition, multiple superflares are detected in the Kepler observations (Figure \ref{Fig.7-flare}), with amplitudes of 4–37 mmag and durations of 46–496 minutes (mean $\approx$ 164 min), exceeding the flare activity of $98\%$ of RS CVn-type binaries reported by \citet{2025A&A...699A.322X}. The flare occurrence rate is ($\sim0.25\%$), corresponding to roughly two events per quarter, though shorter and weaker flares may have been missed. The flare energies range from $\sim 10^{34}$ to $10^{36}$ erg, with most events concentrated around $10^{35}$ erg. The presence of such energetic events indicates strong magnetic activity driven by rapid stellar rotation in this short-period system.

In summary, KIC 6464285 is a typical spectroscopic triple composed of a detached main-sequence binary and a dynamically significant tertiary component, whose presence explains the LTTE-driven orbital-period modulation. The starspot-induced 131-day quasi-periodic variability and the detection of multiple superflares further demonstrate intense magnetic activity within the system. These combined properties make KIC 6464285 an excellent laboratory for studying triple-system dynamics, magnetic activity in close binaries, and the photometric manifestations of rapid stellar rotation.

\begin{acknowledgments}
This work was supported by the International Cooperation Projects of the National Key R$\&$D Program (Grant No. 2022YFE0127300), the National Natural Science Foundation of China (Grant No. 12573038), the International Partnership Program of the Chinese Academy of Sciences (Grant No. 020GJHZ2023030GC), the Yunnan Fundamental Research Projects (Grant Nos. 202503AP140013, 202401AS070046, and 202501AS070055), and the Yunnan Revitalization Talent Support Program.
This work makes use of photometric data from the Kepler and TESS missions obtained from the Mikulski Archive for Space Telescopes (MAST; \citealt{MAST_Kepler, MAST_TESS}), as well as SDSS–APOGEE near-infrared spectroscopy, ZTF public photometry, and LAMOST low-resolution spectra. Complementary ground-based photometry was obtained with the 1-m Cassegrain telescope at Yunnan Observatories and the Zeiss-600 and AZT-22 telescopes at the Maidanak Astronomical Observatory; we are grateful to the staff at these facilities for their assistance during the observations.
\end{acknowledgments}

\bibliography{sample7}{}
\bibliographystyle{aasjournalv7}



\end{document}